\documentclass[9pt,twocolumn,twoside]{opticajnl}
\journal{opticajournal} 

\usepackage{lineno}
\usepackage{xr}
\usepackage{amsmath}

\title{Perfect spatiotemporal optical vortex driven high harmonic generation}

\author[4]{C. Granados}
\author[1,2,3]{B. Kumar Das}
\author[4*]{W. Gao}

\affil[1]{Department of Physics, Guangdong Technion - Israel Institute of Technology, 241 Daxue Road, Shantou, Guangdong, China, 515063}
\affil[2]{Technion - Israel Institute of Technology, Haifa, 32000, Israel}
\affil[3]{Guangdong Provincial Key Laboratory of Materials and Technologies for Energy Conversion, Guangdong Technion - Israel Institute of Technology, 241 Daxue Road, Shantou, Guangdong, China, 515063}
\affil[4]{Eastern Institute of Technology, Ningbo, 315200, China}
\affil[*]{wgao@eitech.edu.cn}

\begin{abstract}
The generation of high-order harmonic beams carrying orbital angular momentum (OAM) promises application in diverse research fields. Recently, the perfect spatiotemporal optical vortex (PSTOV) beam has garnered much attention due to its topological charge (TC)-independent ring size and intensity distribution in the spatiotemporal plane. Here, we theoretically investigate the harmonic generation process in atomic gases driven PSTOV beams carrying transverse OAM. The unique spatiotemporal characteristics of the PSTOV beam make it an excellent candidate to generate harmonic beams with high TC values. We demonstrate that the conservation of transverse OAM is strictly followed when the harmonic generation is driven by the PSTOV beam. Additionally, in the near-field, the intensity distributions of harmonics show tilted lobed structures, which encapsulate information about the TC. By solving the Fraunhofer diffraction integral, we demonstrate that the generated harmonic vortices exhibit similar divergence properties at the far-field. Our research provides a unique route to create Bessel-Gauss spatiotemporal optical vortex beams with high TC in the extreme-ultraviolet (XUV) spectral regime.
\end{abstract}

\setboolean{displaycopyright}{false} 

\begin{document}

\maketitle

Structured light has revolutionized the way we understand the light-matter interaction \cite{ControlOAM} not only from the fundamental point of view \cite{allen1992orbital}, but also from the applications perspective \cite{RMOAM}. Large attention has been devoted to the properties of light beams carrying orbital angular momentum (OAM) along the propagation direction, called spatial or longitudinal optical vortices (OVs) \cite{PM1,RMOAM}. The generation and propagation properties of OVs have been extensively studied in the past~\cite{BEIJ:1993,BEIJ:1994,Tovar:2000,ZHANG2022105327}. Furthermore, OVs have been used to investigate nonlinear laser-matter interaction in order to extend its application range~\cite{Bjorkholm:66,Dholkia:96,Courtial:97,Zhou:14,Lin:13,BKD}. Inducing nonlinear processes using OVs is extremely interesting since it is possible to generate vortices in the extreme-ultraviolet spectral \cite{TSM2} and attosecond temporal regime \cite{AttoVortex}. Different types of OVs were used to drive the high harmonic generation (HHG) process \cite{POVHHG,eHHG,TVOAM,Gauthier19,Delas22} within the framework of the dipole approximation demonstrating that: (1) the OAM is conserved, and (2) different harmonics are emitted with similar divergence. Theoretically, the HHG process was modeled using the thin-slab model (TSM) \cite{TSM2,Hernandez:2015}. This model replaces the original target by a thin 2D slab placed perpendicularly to the beam propagation direction and the slab can be moved inside the Rayleigh range of the focused beam. Additionally, the harmonic field amplitude is expressed as the fundamental field amplitude raised to the power $p$, where $p$ is the scaling factor. Typically, the value of the scaling factor is calculated from the harmonic field amplitude values for different fundamental field amplitudes. Then, the macroscopic response of the target system is calculated in the far-field by solving the Fraunhofer diffraction integral. 

Additionally, it is also possible to generate optical vortex beams which carry OAM in the direction perpendicular to their propagation direction\cite{PKU, STOV_NatPhot,STOV_PRX,STOV2012}. Those OVs are called spatiotemporal optical vortex (STOV) beams. In particular, it was recently demonstrated the possibility to generate experimentally perfect spatiotemporal OVs \cite{POV_STOV}. These beams were also applied to drive the HHG process \cite{STOV_Exp,Li:2025}, demonstrating that the transverse OAM is conserved in the HHG process. However, a little attention has been put into the nonlinear processes driven by STOV beams, its applications and underlying physics. Since the dipole approximation is valid during the vortex light-matter interaction process, new ways to couple the OAM carried by the light into the electron motion are needed. One possible way is the generation of harmonics with very large topological charge (TC) \cite{LargeTC}. However, in the case of conventional vortex beams such as Laguerre-Gaussian or Bessel-Gauss beams, the beam size increases monotonically with the TC, making it difficult to generate beams carrying higher values of the TC\cite{HTC}. This underlying challenge can be tackled easily with the use of perfect optical vortex (POV) beams for HHG \cite{POVHHG}. POV beams are specially designed vortex beams which show TC-independent ring radius, ring width, and annular intensity distribution~\cite{Ostrovsky13,vaity15}. Consequently, it is possible to produce fundamental beams with large TC to drive the harmonic generation process. Recently, the perfect STOV (PSTOV) beams were proposed \cite{PSTOV}, opening the possibility, in the spatiotemporal regime, to generate harmonic fields carrying large TC values. 

In this work, we theoretically investigate the harmonic generation process in atomic gases driven by PSTOV beams. We briefly introduce the PSTOV beam and discuss how various beam features such as the ring size and the ring width alter with changing TCs, at the source plane. Furthermore, we model the interaction of the PSTOV beam with an atomic gas target using the TSM. Moreover, the Fraunhofer diffraction integral is utilized to compute the complex harmonic amplitude at the far-field.


We begin by describing the fundamental PSTOV beam. The field representing the PSTOV beam can be expressed in the $y=0$ plane as \cite{PSTOV}: 
\begin{eqnarray}
    E_{\mathrm{PSTOV}}(x,t)&=& E_0 \exp \left(-i \Phi \right)\exp\left(- \frac{(r_{x\tau}-r_0)^2}{\delta^2 } \right), \nonumber \\
    \label{pstov}
\end{eqnarray}

where $r_{x\tau} = \sqrt{x^2/w_x^2 + \tau^2/w_\tau^2}$, $\tau = c (t - \delta t) - z$, and $\Phi=  - l \phi - \omega_0 t + k_0 z $. The helical phase is given as, $-l\phi = -l\arctan(w_x\tau/w_\tau x)$, where $w_x$ and $w_\tau$ are the spatial and temporal widths, respectively. Furthermore, $r_0$, $\delta$, and $E_{0}$ represent the ring radius, ring thickness, and constant amplitude of the PSTOV beam, respectively. From Eq.~(\ref{pstov}), it can be clearly seen that the spatiotemporal amplitude of the beam is independent of the TC. This holds true only in the asymptotic limit i.e., when $r_0\gg \delta$. Therefore, the ring radius, the ring thickness, and the overall intensity distribution of the PSTOV beam do not change with increasing TC values. However, it is important to highlight that the spatiotemporal amplitude of the beam becomes dependent on the TC in the scenario where the asymptotic condition is not satisfied. In such a scenario: (1) the beam characteristics hardly change for lower values of the TC, (2) significant changes can be observed for extremely higher values of the TC. In the case of spatial POV beams, it is well-known that the radius of the maximum intensity ring, which defines the divergence of the POV beam, is independent of the TC \cite{POVHHG}. Similarly, it can also be shown that the radius of maximum intensity, $r^{max}_{x\tau}$, is also independent of the TC for the PSTOV beam. 

\begin{figure}[h!]
\centering 
\includegraphics[width=0.9\linewidth]{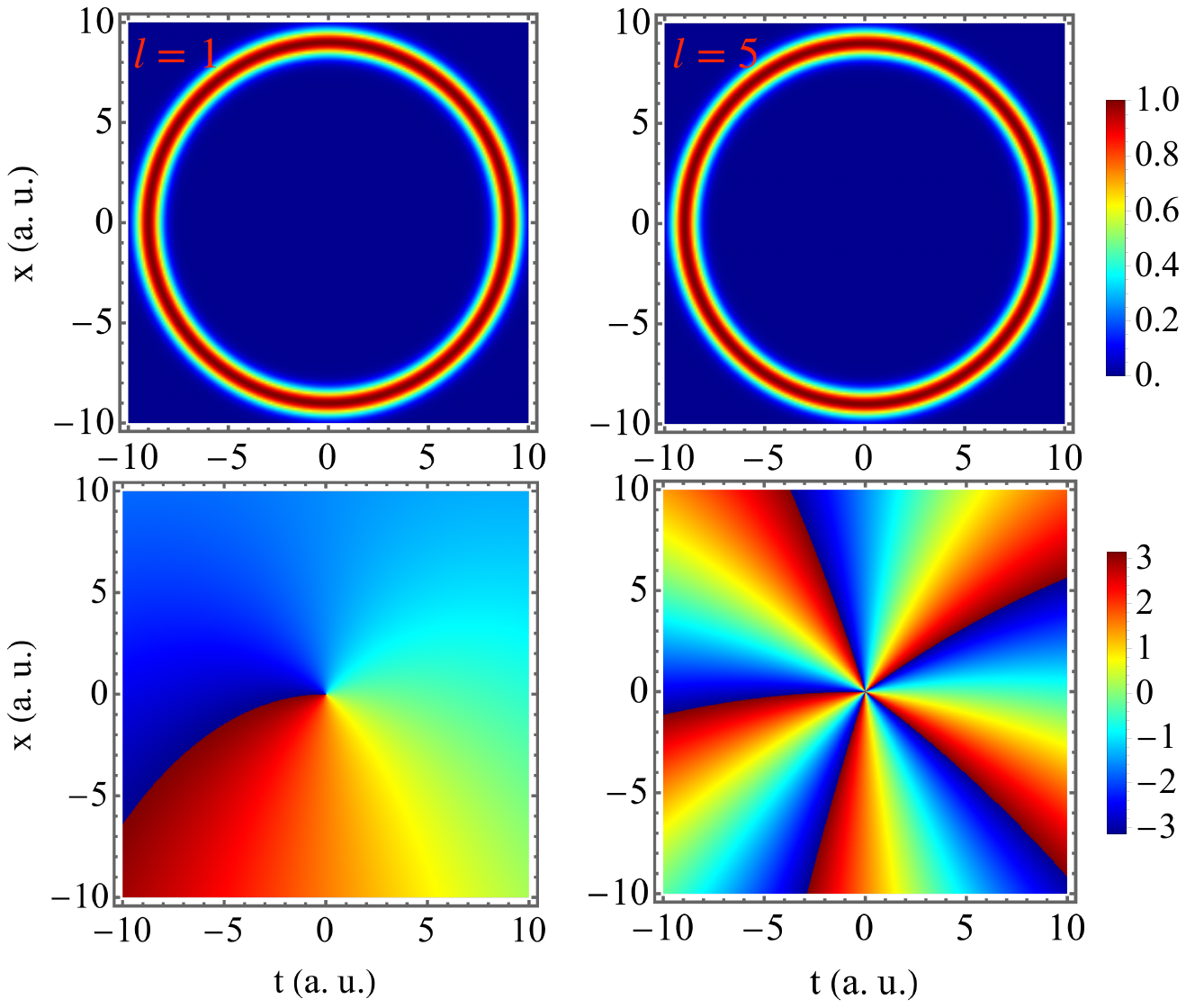}
\caption{Intensity and phase distributions of the fundamental PSTOV beam. In the upper panels, we show the intensity distributions of the PSTOV beam with $l=1$ and 5. In the lower panels, we show their corresponding phase profiles.}
\label{Fig1}
\end{figure}

In Fig.~\ref{Fig1}, we show the intensity (upper panels) and phase (lower panels) distributions of the PSTOV beam for different TC values at the source plane i.e., at $z=0$. For the simulation, we use: $r_{0}=3$~a.u., $\delta=0.3$~a.u., $w_{x}=1$ a.u., $w_{\tau}=411$ a.u. It can be seen from the figure that there is no change in the ring radius and ring width of the beam for TC values $l=1$ and 5. Furthermore, TC information can be extracted by counting the number of $2\pi$ phase shifts along the azimuthal coordinate of the beam. Note that the curved lines, in the phase plot for $l=1$, emerge from the difference between the spatial and temporal widths. 

\begin{figure}[h!]
\centering 
\includegraphics[width=0.9\linewidth]{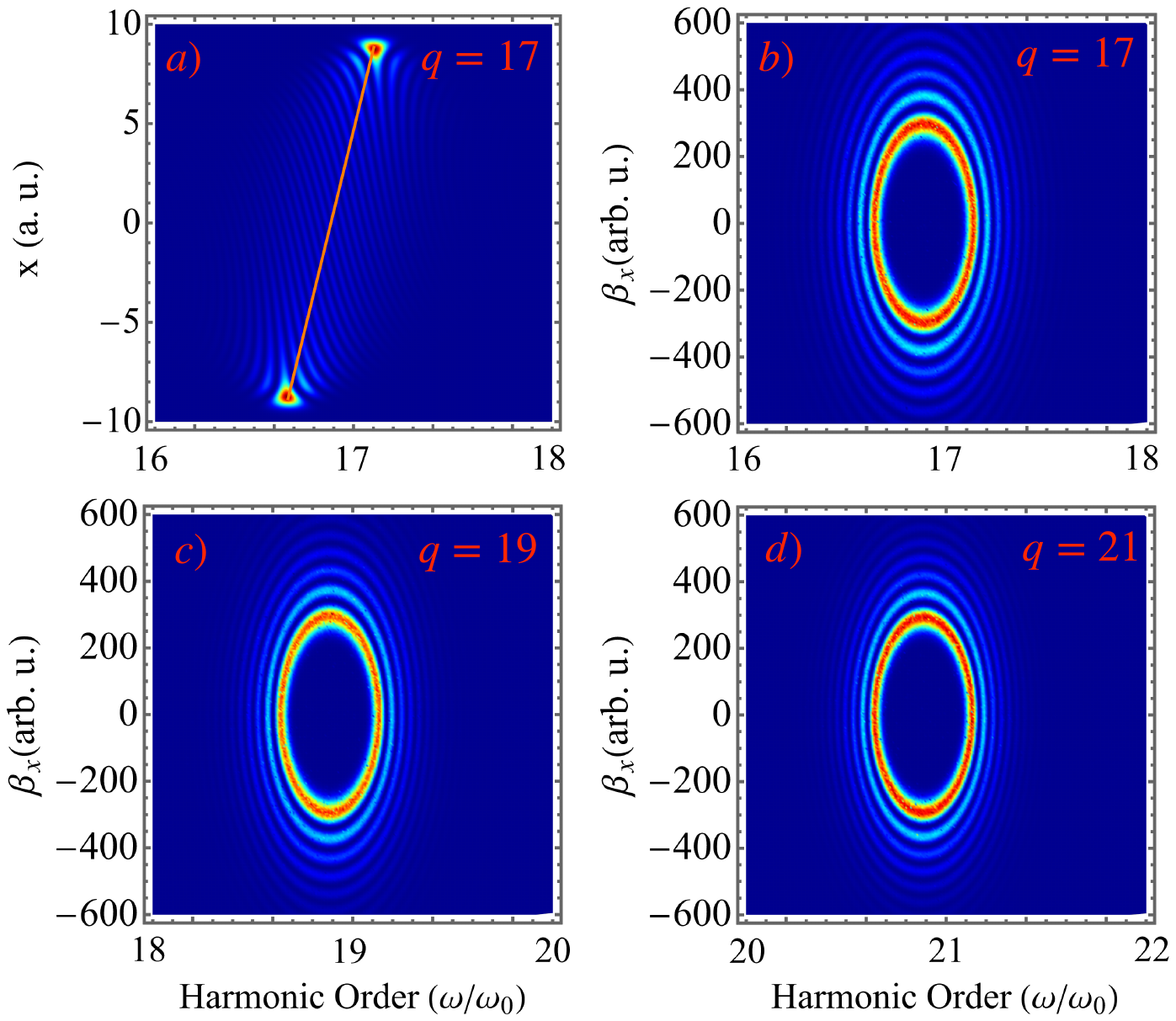}
\caption{Perfect spatiotemporal harmonic vortices. In (a), we show the intensity distribution of the harmonic order 17$^{\text{th}}$ in the near-field. In (b) to (d), we present the intensity distributions of harmonic orders 17, 19, and 21 at the far- field. Here, we use a PSTOV beam with a TC value, $l=1$.}
\label{Fig2}
\end{figure}

In the following we will discuss the harmonic generation process driven by a perfect spatiotemporal vortex beam. We will present first a short description of the theoretical model. To understand the nonlinear interaction of the PSTOV beam with an atomic gas target and the generation of harmonic vortices, we employ the well-known TSM. In this model, the original atomic target is replaced by a thin 2D-slab placed perpendicularly to the propagation direction of the driving beam. Due to the thinness of the slab and transverse structure of the driving beam, longitudinal phase matching is often neglected and transverse phase matching is taken into consideration. Moreover, the slab can be moved and positioned at various locations with respect to the driving beam's focus. In the TSM, the harmonic amplitude is expressed as the driving field's amplitude raised to the power of $p$, where $p$ is the scaling factor. Furthermore, the harmonic phase is expressed as the harmonic order times the driving field's phase plus the dipole phase: 

\begin{eqnarray}
    E_{\mathrm{PSTOV}}^{near}(x,t)&=& E_0^p \exp\left(- \frac{p(r_{x\tau}-r_0)^2}{\delta^2 } \right) \exp\left(-i\Psi_q \right), \label{near}
\end{eqnarray}

where, $\Psi_q = q\Phi-\alpha_qI_{PSTOV}$ represents the total harmonic phase and $\alpha_qI_{PSTOV}$ is the intrinsic dipole phase for the harmonic order, $q$. From Eq.~(\ref{pstov}) and Fig.~\ref{Fig1}, it can be found that there is a thin ring formation at $r_{x\tau}=r_{0}$ and the intensity distribution is uniform unlike the case of Gaussian STOV beams where the intensity distribution is more dispersed. Therefore, for simplicity, we can assume that the harmonic generation takes place for the maximum of intensity. This is also support by the high non-linearity of the HHG process, which dictates that the generation efficiency decreases rapidly outside the maximum of the driving field. It has also been demonstrated that the dipole phase contribution gives rise to secondary OAM which are almost an order of magnitude weaker than that of the helical phase contribution. For these reasons, we neglect the contribution of the dipole phase in our model. The value of the scaling factor,$p$, for harmonics in the spectral plateau is $p\approx3$ and corresponds to a scaling law extracted by varying the driving field's intensity in the range $I =0.87 \text{ to }1.7\times10^{14}$ W/cm$^2$. By calculating the Fourier transform of Eq.~(\ref{near}), the resulting harmonic field lies in the spatiospectral domain, where tilted lobed structures carrying information about the TC of the generated harmonics are expected. To compute the harmonic amplitude and phase at the far-field, we use the Fraunhofer diffraction integral. It is important to highlight that we solve the far-field integral in the space-time ($x'-t'$) domain in order to have the resulting harmonic field in the divergence-frequency ($\beta_{x}-\omega$) domain. The far-field integral reads: 

\begin{eqnarray}
    E_{\mathrm{PSTOV}}^{far}(\beta_x,\omega)&\propto& \int\int dx' dt' E_{\mathrm{PSTOV}}^{near}(x',\tau')  \nonumber \\ &\times&\exp{\Bigg(-ik_q\beta_xx'-iq\omega \tau'\Bigg)} \label{far}
\end{eqnarray}

which is customary for the HHG calculations \cite{STOV_Exp}. In Fig.~\ref{Fig2} (a), we show the near-field intensity distribution of the $17^{th}$ harmonic in the spatiospectral domain. Furthermore, in Fig.~\ref{Fig2}(b)-(d), we show the far-field intensity distributions for harmonic orders 17$^{\text{th}}$ to 21$^{\text{st}}$, respectively in the divergence-frequency domain. For carrying out the numerical simulation, we use the following parameters: $\lambda=800$~nm, $l=1$, $r_{0}=3$ a.u., $\delta=0.3$ a.u., $w_{x}=1$ a.u., $w_{\tau}=411$ a.u. From Fig.~\ref{Fig2}, it can be seen that: (1) the horizontal and vertical radii of the maximum intensity ring are similar for harmonic orders 17,19, and 21, (2) less intense secondary rings are also present in the far-field intensity distributions of those harmonic orders. From (1), it can be concluded that different harmonics are emitted with similar divergence at the far-field. It is important to note that the generation of high-order harmonics with similar radius of the maximum intensity has also been reported for the POV beam driven HHG \cite{POVHHG}. It can be considered as a consequence of the conservation of the OAM in the HHG process. Similarly, the presence of less intense secondary rings in the far-field intensity distributions is a consequence of the free-space propagation of the generated harmonics. This is due to the fact that perfect optical vortex beams and Bessel-Gauss vortex beams are Fourier pairs both in the spatial and spatiotemporal domain~\cite{Bikash2024, Bessel}. It can also be seen from Fig.~\ref{Fig2}(b)-(d) that the total number of intensity rings are the same for all the harmonic orders. This is due to the fact that for a fixed value of $\delta$, the number of rings in the far-field intensity distribution is solely determined by $r_{0}$. Since we have used the same value of $r_{0}$ in all the cases, same number of rings appears in all the plots. Such features have also been demonstrated in the free-space propagation of spatial POV beams. This demonstrates the generation of Bessel-Gauss spatiotemporal optical vortex (Be-STOV) in the far-field. 

It is important to highlight that we also investigated the generation of BeSTOV beams in the perturbative regime i.e., the case of low-order harmonics.  For the low harmonics, we also observed the generation of Be-STOV (no shown here) for the perturbative region.  Additionally, the lobed structure exhibited by the PSTOV beam in the spatiospectral regime, as shown in Fig.~\ref{Fig2} (a), is a characteristic of the conjugate nature of the spatiotemporal beams and the Hermite polynomials. As described in Ref.~\cite{PorrasSTOV}, the resulting number of minima between the primary lobes reflects the TC of the resulting beam (for the $q^{th}$ order harmonic, $q$ number of minima will appear between the primary lobes). In this case, the TC of the vortex harmonic scales linearly with the harmonic order, $l_q=ql$ \cite{SHG_STOV} i.e., $l_{17}=17$ (see Fig.~\ref{Fig2}(a)). This is a manifestation of the conservation of the OAM in spatiotemporal vortex beams driven HHG. 

\begin{figure}[h!]
\centering 
\includegraphics[width=0.8\linewidth]{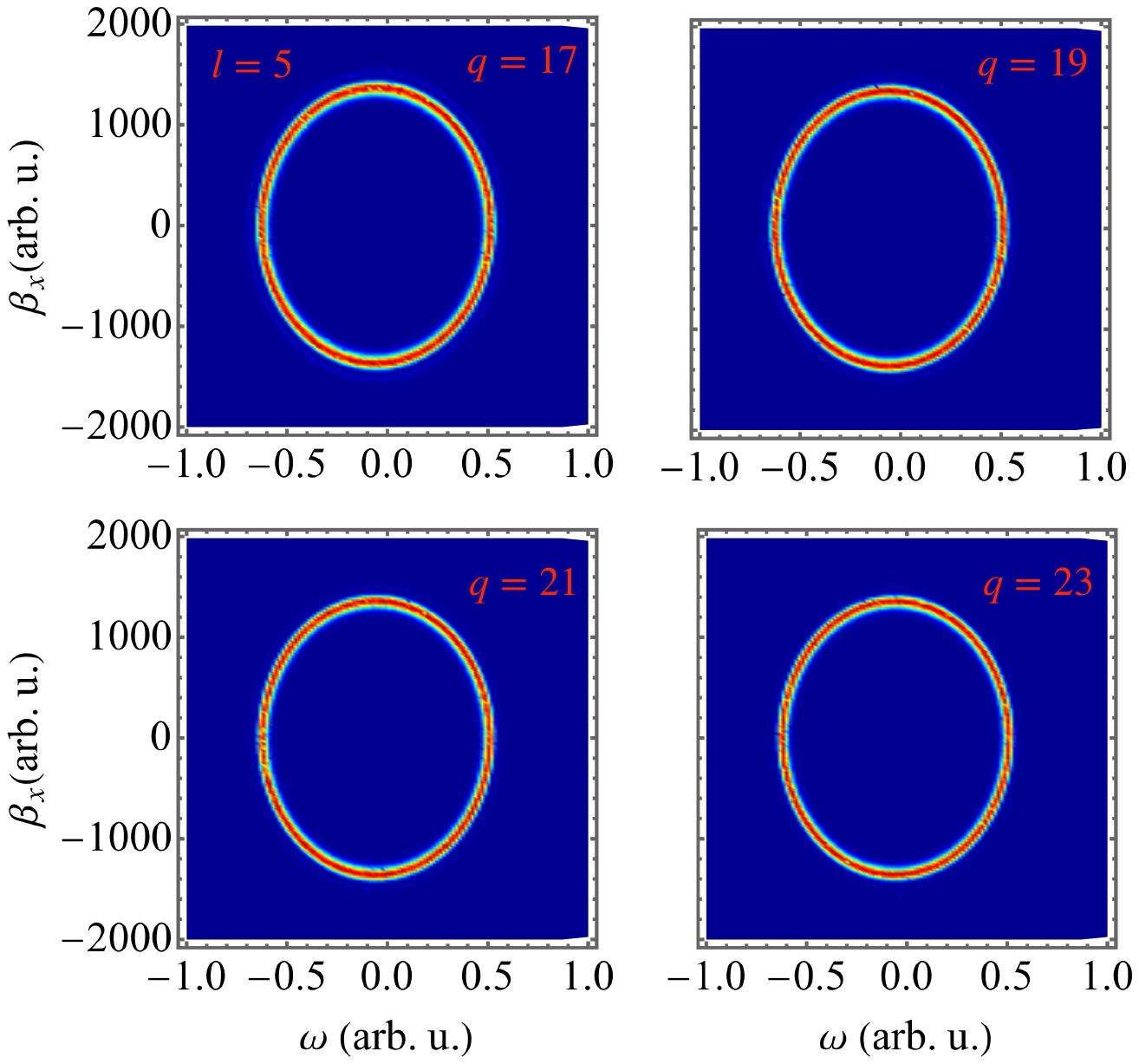}
\caption{Perfect spatiotemporal harmonic vortices. In the different panels, we present the harmonic orders from 17$^{\text{th}}$ to 23$^{\text{rd}}$ generated by a fundamental PSTOV beam with TC value, $l=5$. }
\label{Fig3}
\end{figure}

An important difference between the spatial vortex and the STOV beams is that spatiotemporal couplings (STCs) occur naturally in the STOV beams. Notice that, in the case of STOV beams, the helical phase comprises of both space and time coordinates, $\arctan(w_x\tau/w_\tau x)$, unlike the spatial vortex beams helical phase that contains transverse spatial coordinates. The STCs do not allow the separation of space and time coordinates in an electromagnetic field i.e., $E_{PS}(x,t)\ne E_{PS}(x)E_{PS}(t)$. This makes the STOV beams the quintessential example of STCs. In the harmonic generation, the STCs manifest in the Fourier transform of the beam from the spatiotemporal domain to the spatiospectral (or divergence-time) domain. As shown in Fig.~\ref{Fig2} (a), there is a spatial chirp in the resulting intensity distribution in the spatiospectral domain. The tilt of the beam is a consequence of the STCs and can be described by the following function:

\begin{equation}
    x_{P}(\omega_q)=\frac{n r_0 w_x  w_\tau}{ \delta l  c} \omega_q, \label{STC}
\end{equation}

where $n$ is a number that depends on the type of STOV beam and the value of TC. For our case, $n=0.9$. The central frequency of the harmonic is represented by $\omega_q$. The function $x_{P}(\omega_q)$ represents the slope of the intensity distribution and is represented by the red line in Fig.~\ref{Fig2}(a). Essentially, Eq.~(\ref{STC}) links the energy conservation with the STCs. This can be understood in the following way: in Eq.~(\ref{far}), the conservation of energy in the harmonic field i.e., $\omega_q=q\omega$, is imposed by using the harmonic energy $w_q$ for each harmonic field. In this way, the same chirp function $x_P(\omega)$ is independent of the harmonic order. If one uses $\omega$ instead of $\omega_q$ in Eq.~\ref{far}, the resulting harmonic spectra in the near-field would give rise to a chirp that changes with the harmonic order- a clear violation of the energy conservation principle. Finally, notice that for other type of STOV beams, the chirp function, Eq.~(\ref{STC}), changes accordingly to the Fourier transform of the beam to the spatiospectral domain. 
\\
\\
We further investigate the influence of the TC of the fundamental beam on the far-field harmonic intensity distribution. For this numerical analysis, we use $l=5$ while keeping all other simulation parameters fixed. In Fig.~\ref{Fig3}, we show the far-field intensity distributions of the harmonic orders 17$^\text{th}$, 19$^\text{th}$, 21$^\text{st}$, and 23$^\text{rd}$. It can be seen from the figure that the ring radii as well as the ring width of the generated harmonics hardly change, thereby, indicating the generation of harmonics with similar divergence. A comparison between the two cases i.e., $l=1$ and $l=5$ reveals the following: (1) more external rings are observed in the harmonic intensity distributions for $l=1$ case, whereas, the number of external rings reduces drastically for the $l=5$ case. For instance, the $17^{\text{th}}$ and $19^{\text{th}}$ harmonics show only one external ring apart from the maximum intensity ring (see the upper panels in Fig.~\ref{Fig3}). Furthermore, no external rings are observed in the intensity distributions of the $21^{\text{st}}$ and $23^{\text{rd}}$ harmonics (see the lower panels in Fig.~\ref{Fig3}). (2) In the case of PSTOV beam with $l=5$, the vertical ring radius of the resulting harmonics increases around 4 times than that of the harmonics generated by the PSTOV beam with $l=1$. This is expected, since harmonic vortex intensity in far-field depends on the TC \cite{POVHHG}. In addition, the shape of the harmonic vortex does not change, meaning the small ring size is maintain, however this time, the differences in the spatial and temporal width are enhanced because of the Fourier transform (notice the different thickness for the plane $\beta_x=0$ and $\omega/\omega_0=0$).

In conclusion, we demonstrated the generation of low and high-order harmonics from an atomic gas target using PSTOV beams. To better understand the transfer of vorticity or OAM from the fundamental PSTOV beam to the generated harmonics, we employed the thin-slab model in combination with the Fraunhofer diffraction integral to compute the near and far-field complex harmonic amplitudes. We found that: (1) when the PSTOV beam carrying a TC $l=1$ drives the HHG process, different harmonics are emitted with similar divergence and show BeSTOV intensity profiles. This was not directly observed for the spatial POV driven HHG. (2) When the TC of the fundamental PSTOV beam is increased from $l=1$ to $l=5$, still different harmonics are emitted with similar divergence. However, the number of external rings in the far-field intensity distribution decreases significantly. Additionally, we also observe that for larger harmonic orders, the external ring's intensity decreases drastically. (3) The OAM conservation holds true following a similar OAM upscaling rule i.e., $l_{q}=ql$, where $l_{q}$, and $l$ denote the transverse OAM value of the $q^{th}$ order harmonic, and the fundamental beam, respectively. (4) Since the STCs naturally appear in the STOV beams, we discussed its manifestation in the near-field and demonstrated that it can be linked to the conservation of energy in the harmonic field. Only when the energy conservation principle is properly accounted for, the resulting spatial chirp becomes independent of the harmonic order. Our work opens the door to the generation of BeSTOV beams in the ultraviolet and extreme-ultra violet spectral regimes. Added to this, our investigation can be potentially be applied to further understand the interaction of vortex beams with matter and its effects beyond the dipole approximation. Since various high-order harmonics are emitted with similar divergence at the far-field, they will have a good spatial overlap and focusability for practical applications.  Furthermore, these far-field properties can be used to generate spatiotemporal light springs. Our research can also be extended to generate STOV beams in solid targets and investigate the coupling of the electromagnetic field's topology and the topology of the solid materials. This is an important open question that can be solved by means of structured light in the UV and XUV spectral regimes. As a future outlook, it would be interesting to exploit the interplay between the spin and orbital angular momentum of PSTOV beams in a non-collinear geometry to generate spatiotemporal attosecond pulse train.







\begin{backmatter}

\bmsection{Disclosures} The authors declare no conflicts of interest.
\end{backmatter}

\bibliography{sample}

\begin{thebibliography}{10}
\newcommand{\enquote}[1]{``#1''}

\bibitem{ControlOAM}
J.-L. Bégin, E.~Karimi, P.~Corkum, \emph{et~al.},
  {\protect\JournalTitle{Nature Communications}} \textbf{16}, 2467 (2025).

\bibitem{allen1992orbital}
L.~Allen, M.~Beijersbergen, R.~Spreeuw, and J.~Woerdman,
  {\protect\JournalTitle{Phys. Rev. A}} \textbf{45}, 8185 (1992).

\bibitem{RMOAM}
A.~Forbes, M.~de~Oliveira, and M.~R. Dennis, {\protect\JournalTitle{Nature
  Photonics}} \textbf{15}, 253 (2021).

\bibitem{PM1}
Y.~Shen, X.~Wang, Z.~Xie, \emph{et~al.}, {\protect\JournalTitle{Light: Science
  \& Applications}} \textbf{8}, 90 (2019).

\bibitem{BEIJ:1993}
M.~Beijersbergen, L.~Allen, H.~{van der Veen}, and J.~Woerdman,
  {\protect\JournalTitle{Optics Communications}} \textbf{96}, 123 (1993).

\bibitem{BEIJ:1994}
M.~Beijersbergen, R.~Coerwinkel, M.~Kristensen, and J.~Woerdman,
  {\protect\JournalTitle{Optics Communications}} \textbf{112}, 321 (1994).

\bibitem{Tovar:2000}
A.~A. Tovar, {\protect\JournalTitle{J. Opt. Soc. Am. A}} \textbf{17}, 2010
  (2000).

\bibitem{ZHANG2022105327}
Y.~Zhang, C.~Ke, Y.~Xie, and Y.~Zhang, {\protect\JournalTitle{Results in
  Physics}} \textbf{35}, 105327 (2022).

\bibitem{Bjorkholm:66}
J.~E. Bjorkholm, {\protect\JournalTitle{Phys. Rev.}} \textbf{142}, 126 (1966).

\bibitem{Dholkia:96}
K.~Dholakia, N.~B. Simpson, M.~J. Padgett, and L.~Allen,
  {\protect\JournalTitle{Phys. Rev. A}} \textbf{54}, R3742 (1996).

\bibitem{Courtial:97}
J.~Courtial, K.~Dholakia, L.~Allen, and M.~J. Padgett,
  {\protect\JournalTitle{Phys. Rev. A}} \textbf{56}, 4193 (1997).

\bibitem{Zhou:14}
Z.-Y. Zhou, Y.~Li, D.-S. Ding, \emph{et~al.}, {\protect\JournalTitle{Opt.
  Express}} \textbf{22}, 23673 (2014).

\bibitem{Lin:13}
Y.~C. Lin, K.~F. Huang, and Y.~F. Chen, {\protect\JournalTitle{Laser Phys.}}
  \textbf{23}, 115405 (2013).

\bibitem{BKD}
B.~K. Das, M.~Kr{\"u}ger, and M.~F. Ciappina, \enquote{{Second-harmonic
  generation in a KTP crystal driven by vortex beams with integer topological
  charges},}  (SPIE, 2025), p. 1354106.

\bibitem{TSM2}
C.~Hern\'andez-Garc\'{\i}a, A.~Pic\'on, J.~San~Rom\'an, and L.~Plaja,
  {\protect\JournalTitle{Phys. Rev. Lett.}} \textbf{111}, 083602 (2013).

\bibitem{AttoVortex}
A.~de~las Heras, D.~Schmidt, J.~S. Rom\'{a}n, \emph{et~al.},
  {\protect\JournalTitle{Optica}} \textbf{11}, 1085 (2024).

\bibitem{POVHHG}
B.~K. Das, C.~Granados, M.~Kr\"uger, and M.~F. Ciappina,
  {\protect\JournalTitle{Phys. Rev. Res.}} \textbf{6}, 043244 (2024).

\bibitem{eHHG}
C.~Granados, B.~K. Das, W.~Gao, and M.~F. Ciappina,
  {\protect\JournalTitle{Applied Physics Letters}} \textbf{126}, 081104 (2025).

\bibitem{TVOAM}
L.~Rego, K.~M. Dorney, N.~J. Brooks, \emph{et~al.},
  {\protect\JournalTitle{Science}} \textbf{364}, eaaw9486 (2019).

\bibitem{Gauthier19}
D.~Gauthier, S.~Kaassamani, D.~Franz, \emph{et~al.},
  {\protect\JournalTitle{Opt. Lett.}} \textbf{44}, 546 (2019).

\bibitem{Delas22}
A.~de~las Heras, A.~K. Pandey, J.~S. Rom\'{a}n, \emph{et~al.},
  {\protect\JournalTitle{Optica}} \textbf{9}, 71 (2022).

\bibitem{Hernandez:2015}
C.~Hernández-García, J.~S. Román, L.~Plaja, and A.~Picón,
  {\protect\JournalTitle{New J. Phys.}} \textbf{17}, 093029 (2015).

\bibitem{PKU}
Y.~Fang, S.~Lu, and Y.~Liu, {\protect\JournalTitle{Phys. Rev. Lett.}}
  \textbf{127}, 273901 (2021).

\bibitem{STOV_NatPhot}
A.~Chong, C.~Wan, J.~Chen, and Q.~Zhan, {\protect\JournalTitle{Nat. Photonics}}
  \textbf{14}, 350 (2020).

\bibitem{STOV_PRX}
N.~Jhajj, I.~Larkin, E.~W. Rosenthal, \emph{et~al.},
  {\protect\JournalTitle{Phys. Rev. X}} \textbf{6}, 031037 (2017).

\bibitem{STOV2012}
K.~Y. Bliokh and F.~Nori, {\protect\JournalTitle{Phys. Rev. A}} \textbf{86},
  033824 (2012).

\bibitem{POV_STOV}
H.~Fan, Q.~Cao, X.~Liu, \emph{et~al.}, {\protect\JournalTitle{Photon. Res.}}
  \textbf{13}, 1776 (2025).

\bibitem{STOV_Exp}
R.~Martín-Hernández, G.~Gui, L.~Plaja, \emph{et~al.},
  {\protect\JournalTitle{Nature Photonics}}  (2025).

\bibitem{Li:2025}
X.~Li, X.~Zhang, X.~Zheng, \emph{et~al.}, {\protect\JournalTitle{New Journal of
  Physics}} \textbf{27}, 023008 (2025).

\bibitem{LargeTC}
W.~Chen, W.~Zhang, Y.~Liu, \emph{et~al.}, {\protect\JournalTitle{Nature
  Communications}} \textbf{13}, 4021 (2022).

\bibitem{HTC}
A.~K. Pandey, A.~de~las Heras, T.~Larrieu, \emph{et~al.},
  {\protect\JournalTitle{ACS Photonics}} \textbf{9}, 944 (2022). Publisher:
  American Chemical Society.

\bibitem{Ostrovsky13}
A.~S. Ostrovsky, C.~Rickenstorff-Parrao, and V.~Arriz\'{o}n,
  {\protect\JournalTitle{Opt. Lett.}} \textbf{38}, 534 (2013).

\bibitem{vaity15}
P.~Vaity and L.~Rusch, {\protect\JournalTitle{Opt. Lett.}} \textbf{40}, 597
  (2015).

\bibitem{PSTOV}
S.~A. Ponomarenko and D.~Hebri, {\protect\JournalTitle{Opt. Lett.}}
  \textbf{49}, 4322 (2024).

\bibitem{Bikash2024}
B.~K. Das, C.~Granados, and M.~F. Ciappina, {\protect\JournalTitle{Appl. Opt.}}
  \textbf{63}, 2737 (2024).

\bibitem{Bessel}
A.~S. Rao, {\protect\JournalTitle{Physica Scripta}} \textbf{99}, 062007 (2024).

\bibitem{PorrasSTOV}
M.~A. Porras, {\protect\JournalTitle{Nanophotonics}}  (2025).

\bibitem{SHG_STOV}
G.~Gui, N.~J. Brooks, H.~C. Kapteyn, \emph{et~al.},
  {\protect\JournalTitle{Nature Photonics}} \textbf{15}, 608 (2021).

\end{thebibliography}
\bibliographyfullrefs{sample}




\end{document}